\documentclass[review]{elsarticle}

\usepackage{lineno,hyperref}
\modulolinenumbers[5]

\journal{Journal of \LaTeX\ Templates}

\usepackage{bm}
\usepackage{mathrsfs}
\usepackage[english]{babel}
\usepackage{amsmath,cases,amsfonts,amssymb}
\usepackage{epsfig, multirow, subfigure, txfonts}
\usepackage[small]{caption2}
\usepackage{color,hyperref}
\usepackage[thmmarks]{ntheorem}
{
\theoremstyle{nonumberplain}
\theoremheaderfont{\bfseries}
\theorembodyfont{\normalfont}
\theoremsymbol{\mbox{$\Box$}}
\newtheorem{Proof}{Proof.}
}

\def\bc{\begin{center}}
      \def\ec{\end{center}}

 \def\bel{\begin{equation}}
 \def\enl{\end{equation}}
 \def\be{\begin{eqnarray*}}
 \def\en{\end{eqnarray*}}
 \def\R{{\bf R}}

\def\i{{\bf i}}
\def\j{{\bf j}}
\def\k{{\bf k}}
 \def\H{{\cal H}}

 \def\ux{\underline{x}}
 \def\uq{\underline{q}}
 \def\uxi{\underline{\xi}}
 \def\ut{\underline{t}}
 
\def\up{\underline{p}}

\newtheorem{Th}{Theorem}[section]
 \newtheorem{Lem}{Lemma}[section]
\newtheorem{Cor}{Corollary}[section]

\newtheorem{Def}{Definition}[section]
\newtheorem{Rem}{Remark}[section]
\newtheorem{Exa}{Example}[section]
\def\i{{\bf i}}

 \def\en{{\bf e}_n}
 
 \def\j{{\bf j}}
 \def\k{{\bf k}}
 
 \def\ux{\underline{x}}

\def\uxi{\underline{\xi}}
 \def\ut{\underline{t}}

 \def\uq{\underline{q}}

\date{}

\begin{document}

\begin{frontmatter}

\title{Uncertainty Principle for Measurable Sets and Signal Recovery in Quaternion Domains}
\author[firstaddress]{Kit Ian Kou}

\author[mymainaddress]{Yan Yang\corref{mycorrespondingauthor}}
\cortext[mycorrespondingauthor]{Corresponding author}
\ead{mathyy@sina.com}

\author[thirdaddress]{Cuiming Zou}


\address[firstaddress]{Department of Mathematics, Faculty of Science and Technology, University of Macau, Taipa, Macao, China. Email: kikou@umac.mo, }
\address[mymainaddress]{School of Mathematics (Zhuhai), Sun Yat-Sen University, Zhuhai, China.}
\address[thirdaddress]{Department of Mathematics, Faculty of Science and Technology, University of Macau, Taipa, Macao, China. Email: zoucuiming2006@163.com}


\begin{abstract}
The classical uncertainty principle of harmonic analysis states that a nontrivial
function and its Fourier transform cannot both be sharply localized.
It plays an important role in signal processing and physics. This paper generalizes the uncertainty principle for measurable sets from complex domain to hypercomplex domain using quaternion algebras, associated with the Quaternion Fourier transform. The performance is then evaluated in signal recovery problems where there is an interplay of missing and
time-limiting data.
\end{abstract}



\end{frontmatter}


\section{Introduction}
The classical uncertainty principle (the continuous-time uncertainty principle)
states that if a function $f(t)$ is essentially zero outside an interval of length $\Delta_t$ and its Fourier transform $\hat{f}(\omega)$ (defined by
($\hat{f}(\omega)=\int_{-\infty}^{\infty}f(t)e^{-2\pi\i t\omega}dt$)
is essentially zero outside an interval of length $\Delta_{\omega}$, then
\begin{eqnarray}\label{yk4eq1}
\Delta_t\Delta_{\omega}\geq 1.
\end{eqnarray}
That means a function and its Fourier transform cannot both be higher concentrated. It was recently generalized from intervals to measurable sets \cite{DS1989}. If $f(t)$ is practically zero outside a measurable set $T$ and $\hat{f}(\omega)$ is practically zero outside a measurable set $W$, then
\begin{eqnarray}\label{yk4eq2}
|T||W|\geq 1-\delta,
\end{eqnarray}
where $|t|$ and $|W|$ denote the measures of the sets $T$ and $W$, and $\delta$ is a small number bound.

The quaternion Fourier transform (QFT) plays a vital role in the representation of (hypercomplex) signals.
It transforms a real (or quaternionic) 2D signal into a quaternion-valued
frequency domain signal. The four components of the QFT separate four
cases of symmetry into real signals instead of only two as in the complex
FT. In \cite{B1999, SE2007} the authors used the QFT to process color image analysis. The
paper \cite{BLC2003} implemented the QFT to design a color image digital watermarking
scheme. The authors in \cite{BTN2007} applied the QFT to image pre-processing and
neural computing techniques for speech recognition. Recently, certain asymptotic
properties of the QFT were analyzed and a straightforward generalization
of the classical Bochner-Minlos theorem to the framework of quaternion
analysis was derived \cite{GMKS2013}. In this paper, we study the uncertainty principle of measurable sets (\ref{yk4eq2}) associated with QFT,
 the generalization of the 2D Fourier transform (FT) in the Hamiltonian quaternion
algebra. The main motivation of the present study is to develop
further iterative methods for signal recovery problems and to investigate
the corresponding problems in quaternion analysis setting. Further investigations and extensions of this topic will be
reported in a forthcoming paper.

The article is organized as follows. Section \ref{S2} gives a brief introduction
to some general definitions and basic properties of quaternion analysis. The uncertainty principle for measurable sets is
generalized for the right-sided Quaternion Fourier transform of
quaternion-valued signals in Section \ref{S3}. In Section \ref{S4}, applications to signal recovery problems were studied, which can be used to recover
a bandlimited hypercomplex signal with missing data. The proposed algorithm for hypercomplex signal recovery problem are given. We test the performance of the proposed algorithm on two different size of Lena and Chillies images. Moreover we compare their performances. Experimental results demonstrate the advantages of the proposed algorithm in Section \ref{S5}.
Some conclusions are drawn in Section \ref{S6}.

\section{Preliminaries}\label{S2}

The quaternion algebra $\cal{H}$ was first invented by W.
R. Hamilton in 1843 for extending complex numbers to a 4D algebra
\cite{S1979}. A quaternion $q\in \cal{H}$ can be written in this
form
$$q=q_0+\uq=q_0+\i q_1+\j q_2+\k q_3, \; q_k\in \R, \; k=0, 1, 2, 3,$$
where $\i, \j, \k$ satisfy Hamilton's multiplication rules
$$\i^2=\j^2=\k^2=-1, \i\j=-\j\i=\k, $$
$$\j\k=-\k\j=\i, \k\i=-\i\k=\j.$$

Using the Hamilton's multiplication rules, the multiplication of two
quaternions $p=p_0+\up$ and $q=q_0+\uq$ can be expressed as
$$pq=p_0q_0+\up\cdot\uq+p_0\uq+q_0 \up+\up\times\uq,$$
where $\up\cdot\uq=-(p_1q_1+p_2q_2+p_3q_3)$ and
$\up\times\uq=\i(p_3q_2-p_2q_3)+\j(p_1q_3-p_3q_1)+\k(p_2q_1-p_1q_2).$

We define the conjugation of $q\in \cal{H}$ by $\overline{q}=q_0-\i
q_1-\j q_2-\k q_3$. Clearly,
$q\bar{q}=q_0^2+q_1^2+q_2^2+q_3^2.$ So the modulus of a quaternion
$q$ defined by
$$|q|=\sqrt{q\bar{q}}=\sqrt{q_0^2+q_1^2+q_2^2+q_3^2}.$$

In this paper, we study the quaternion-valued signal $f:
\R^2\rightarrow \H$ that can be expressed as
$$f(\ux)=f_0(\ux)+\i f_1(\ux)+\j f_2(\ux)+\k f_3(\ux),$$ where $\ux=x_1\i+x_2\j \in \R^2$
and $f_k, k=0, 1, 2, 3$ are real-valued functions.

For $1\leq p < \infty$, the quaternion modules $L^p(\R^2,\H)$ are defined
as
\begin{eqnarray*}
L^p=L^p(\R^2, \H) := \{f|f:\R^2\rightarrow\H,\|f\|^p_{L^p}:=\int_{\R^2}|f(\ux)|^p
d\ux <\infty\}.
\end{eqnarray*}
Let $f\in L^1{(\R^2, \cal{H})}$, the (right-sided) {\it Quaternion Fourier
transform (QFT)} of $f$ is defined by
\begin{eqnarray}\label{QFT}
{\cal F}\{f\}(\uxi) :=\frac{1}{2\pi}\int_{\R^2}f(\ux)e^{-\i x_1\xi_1}e^{-\j x_2\xi_2}d\ux
\end{eqnarray}
and if in addition, ${\cal F}\{f\} \in L^1\bigcap L^2(\R^2,\H)$, function $f$ can be recovered by its QFT as
$$f(\ux)=\frac{1}{2\pi}\int_{\R^2}{\cal F}\{f\}(\uxi)e^{\j x_2\xi_2}e^{\i x_1\xi_1}d\uxi.$$

The inner product of $f(\ux), g(\ux)\in L^2(\R^2, \cal{H})$ is
defined by
$$<f(\ux), g(\ux)> :=\int_{\R^2}f(\ux)\overline{g(\ux)}d\ux.$$ Clearly,
$\|f\|_{L^2}^2=<f,f>$. In this paper, we consider unit energy signal
for simplification. That is, $\|f\|_{L^2}=1$. By Parseval's identity,
$$\int_{\R^2}|f(\ux)|^2d\ux=\int_{\R^2}|{\cal F}\{f\}(\uxi)|^2d\uxi,$$
we have $\|{\cal F}\{f\}\|_{L^2}=1$ as well. It means that the QFT preserves the energy of the quaternion-valued signal.

\section{Uncertainty Principles}\label{S3}

The uncertainty principle of harmonic analysis states that
a non-trivial function and its FT cannot both
be sharply localized. The uncertainty principle plays an important role
in signal processing \cite{HS1986, DC1991, MS2003, LKMD2015, SV2001, M1991,
HLP1951, OA1995, D2002, LC2004, XWX2009, S2007, CKL2015, KOM2013, KOM2016, YK2014, YK2016}, and physics
\cite{M1990, I1985, MU1988, I2006, I2007, R1960, AO1995, C2000,
W1984, S2008}. In quantum mechanics an uncertainty principle asserts that
one cannot be certain of the position and of the
velocity of an electron (or any particle) at the same time. That is, increasing the
knowledge of the position decreases the knowledge of the velocity or
momentum of an electron. In quaternion analysis some  researches combined the uncertainty relations and the QFT
\cite{BHHA2008, H2010, NS1994, YK2016}. In this section we generalize the uncertainty principle for measurable sets associated with QFT. To process, we first define the $\varepsilon-$concentrated on a measurable set in the space and frequency domains.
\begin{Def}
Let $f: \R^2 \rightarrow \H$ be $\varepsilon-$concentrated on a measurable set $T\subseteq\R^2$, if there is a function $g: \R^2 \rightarrow \H$ vanishing outside $T$ such that $$\|f-g\|_{L^2}<\varepsilon.$$
\end{Def}
Similarly,
\begin{Def}
If $f \in L^1(\R^2,\H)$, then its QFT ${\cal F} \{f\}$ is $\varepsilon-$concentrated on a measurable set $W\subseteq\R^2$ if there is a function $h: \R^2 \rightarrow \H$ vanishing outside $W$ with
$$\|{\cal F}\{f\}-h\|_{L^2}<\varepsilon.$$
\end{Def}

Now we state the main result.
\begin{Th}\label{yk4th1}
Let $T$ and $W$ be measurable sets on $\R^2$ and suppose there is a Quaternion Fourier transform pair $\left(f(\ux), {\cal F}\{f\}(\uxi)\right)$, with $f$ and ${\cal F}\{f\}$ of unit norm, such that $f$ is $\varepsilon_{T}-$concentrated on $T$ and ${\cal F}\{f\}$ is $\varepsilon_{W}-$concentrated on $W$. Then we have
$$|T||W|\geq[1-(\varepsilon_{T}+\varepsilon_{W})]^2.$$
Here $|T|$ and $|W|$ are the measures of the sets $T$ and $W$.
\end{Th}

From Theorem \ref{yk4th1},  we can immediately obtain the following corollary.

\begin{Cor}\label{yk4cor1}
Let $T$ and $W$ be measurable sets on $\R^2$ and suppose that there is a Quaternion Fourier transform pairs $\left(f(\ux), {\cal F}\{f\}(\uxi)\right)$, with $f$ and ${\cal F}\{f\}$ of unit norm (energy), such that $f$ and ${\cal F}\{f\}$ are compact supports on the measurable sets $T$ and $W$, respectively. Then we have
$$|T||W|\geq 1.$$
\end{Cor}

Theorem \ref{yk4th1} and Corollary \label{yk4cor1} generalizes the results from the complex case \cite{DS1989} to the quaternion algebra. Before to proceed the proof of Theorem \ref{yk4th1}, we introduce two crucial operators on $f: \R^2 \rightarrow \H$, namely the space-limiting operator
$$(S_{T}f)(\ux) :=\chi_{T}(\ux)f(\ux),$$
where
\begin{eqnarray*}
\chi_T(\ux) :=\left\{
\begin{array}{lll}
f(\ux), & \ux\in T,\\
0,& \ux \not\in T,
\end{array}
\right.
\end{eqnarray*}
and the frequency-limiting operator
$$(F_{W} f)(\ux) :=\frac{1}{2\pi}\int_{W}{\cal F}\{f\}(\underline{\omega})e^{\j x_2\omega_2}e^{\i x_1\omega_1}d\underline{\omega}.$$
Clearly, we have ${\cal{F}}\{F_{W}f\}(\underline{\omega})=\chi_{W}(\underline{\omega}){\cal F}\{f\}(\underline{\omega})$.

For all $f \in L^2(\R^2,\H)$, given the kernel $k: \R^2 \times \R^2 \rightarrow \H$ which satisfies the following two conditions: $f (\cdot)k(\ut, \cdot) \in L^1(\R^2,\H)$ for almost every $\ut \in \R^2$ and if $$Qf(\ux) :=\int_{\R^2}f(\ut)k(\ut, \ux)d\ut,$$ then $Qf \in L^2(\R^2,\H)$.

Then we define the norm of $Q$ to be
$$\|Q\| :={\rm sup}_{f\in L^2}\frac{\|Qf\|_{L^2}}{\|f\|_{L^2}}={\rm sup}_{f\in L^2}{\|Qf\|_{L^2}}$$ for any unit energy $f$
and the Hilbert-Schmidt norm of $Q$ to be
$$\|Q\|_{HS} :=\left(\int_{\R^2}\int_{\R^2}| k(\ut, \ux)|^2d\ut d\ux\right)^{\frac{1}{2}}.$$
Using Cauchy-Schwarz inequality, we can easily obtain the following lemma.
\begin{Lem}\label{lemadd2}
$\|Q\|_{HS}\geq\|Q\|.$
\end{Lem}

To begin the proof of Theorem \ref{yk4th1}, we digress briefly to
make the following observation.
\begin{Lem}\label{lemadd1}
$\|S_{T}F_{W}\|_{HS}=\|F_{W}S_{T}\|_{HS}.$
\end{Lem}

\begin{Proof}
Using the definition of Quaternion Fourier transform (\ref{QFT}), we have
\begin{eqnarray}\label{addeqma1}
(F_{W} S_{T}f)(\underline{x})&=&\frac{1}{2\pi}\int_{W}{\cal{F}}\{S_Tf\}(\underline{\omega})e^{\j x_2\omega_2}e^{\i x_1\omega_1}d\underline{\omega}, \quad \ux \in \R^2\nonumber\\
&=&\frac{1}{(2\pi)^2}\int_{W}\left(\int_{T}f(\ut)e^{-\i t_1\omega_1}e^{-\j t_2\omega_2}d\ut\right)e^{\j x_2\omega_2}e^{\i x_1\omega_1}d\underline{\omega}\nonumber\\
&=&\frac{1}{(2\pi)^2}\int_{T}f(\ut)\left(\int_{W}e^{-\i t_1\omega_1}e^{-\j t_2\omega_2}e^{\j x_2\omega_2}e^{\i x_1\omega_1}d\underline{\omega}\right)d\ut\nonumber\\
&=&\int_{T}f(\ut) k(\underline{t}, \ux)d\ut,
\end{eqnarray}
where
\begin{eqnarray}\label{addeqma2}
k(\ut, \underline{x} ) := \left\{ \begin{array}{ll} \frac{1}{(2\pi)^2}\int_{W}e^{-\i
t_1\omega_1}e^{-\j t_2\omega_2}e^{\j x_2\omega_2}e^{\i
x_1\omega_1}d\underline{\omega}, & \ut \in T \mbox{ and }  \ux \in \R^2,\\
0, & \mbox{ otherwise.}\end{array}\right.
\end{eqnarray}
While
\begin{eqnarray}\label{addeqma3}
(S_{T} F_{W}f)(\underline{x})&=&\chi_{T}(\ux)(F_{W}f)(\underline{x})\nonumber\\
&=&\chi_{T}(\ux)\frac{1}{2\pi}\int_{W}{\cal F}\{f\}(\underline{\omega})e^{\j x_2\omega_2}e^{\i x_1\omega_1}d\underline{\omega}\nonumber\\
&=&\chi_{T}(\ux)\frac{1}{(2\pi)^2}\int_{W}\left(\int_{\R^2}f(\ut)e^{-\i t_1\omega_1}e^{-\j t_2\omega_2}d\ut\right)e^{\j x_2\omega_2}e^{\i x_1\omega_1}d\underline{\omega}\nonumber\\
&=&\chi_{T}(\ux)\frac{1}{(2\pi)^2}\int_{\R^2}f(\ut)\left(\int_{W}e^{-\i t_1\omega_1}e^{-\j t_2\omega_2}e^{\j x_2\omega_2}e^{\i x_1\omega_1}d\underline{\omega}\right)d\ut\nonumber\\
&=&\chi_{T}(\ux)\int_{\R^2}f(\ut)k(\underline{t}, \ux)d\ut.
\end{eqnarray}
From (\ref{addeqma1}) and (\ref{addeqma3}), we have
$$\|F_{W}S_{T}\|_{HS}=\left(\int_{\R^2}\int_{T}|k(\ut, \ux)|^2d\ut d\ux\right)^{\frac{1}{2}}$$
and
$$\|S_{T}F_{W}\|_{HS}=\left(\int_{T}\int_{\R^2}|k(\ut, \ux)|^2d\ut d\ux\right)^{\frac{1}{2}}.$$
From (\ref{addeqma2}), we know that $\overline{k(\ut, \ux)}=k(\ux,
\ut)$. Therefore, we have $|k(\ut, \ux)|^2=|k(\ux, \ut)|^2$. This
completes the proof.
\end{Proof}

\begin{Rem}
From (\ref{addeqma1}) and (\ref{addeqma3}), we found that the product of two
operators $S_T$ and $F_W$ are not commute.  Fortunately, we can
prove that the HS-norms of these operators $S_T$ and $F_W$ are commute, although Quaternion algebra is a non-commutative algebra.
\end{Rem}

From another application of Eq. (\ref{addeqma1}), we have the
following lemma.
\begin{Lem}\label{lemadd3}
$\|F_{W}S_{T}\|_{HS}=\sqrt{|T||W|}.$
\end{Lem}

\begin{Proof}
Applying (\ref{addeqma1}), we have
$$\|F_{W}S_{T}\|_{HS}^2=\|Q\|_{HS}^2=\int_{\R^2}\int_{T}|k(\ut, \ux)|^2d\ut d\ux.$$
Let $g_{\ut}(\ux) :=k(\ut, \ux)$, where $k(\ut, \ux)$ is given by
(\ref{addeqma2}). Note that
${\cal{F}}\{g_{\ut}\}(\underline{\omega})=\chi_{W}(\underline{\omega}) e^{-\i
t_1\omega_1}e^{-\j t_2\omega_2}$. By Parseval's identity, we have
\begin{eqnarray*}
\int_{\R^2}|g_{\ut}(\ux)|^2d\ux&=&\int_{\R^2}|{\cal{F}}\{g_{\ut}\}(\underline{\omega})|^2d\underline{\omega}
=\int_{W} 1 d\underline{\omega}=|W|.
\end{eqnarray*}
Therefore, we have $\|F_{W}S_{T}\|_{HS}^2=|T||W|.$ This completes the proof.
\end{Proof}

Now, we proceed the proof of Theorem \ref{yk4th1}.

\begin{Proof}
 [Proof of Theorem \ref{yk4th1}]
 Consider the operator $F_{W}S_{T}$, by Parseval's equality and the triangle inequality, applying the assumptions of $f$ is $\varepsilon_{T}$-concentrated on $T$ and ${\cal F} \{f\}$ is $\varepsilon_{W}$-concentrated on $W$, we have
\begin{eqnarray}\label{ky1}
\|f-F_{W}S_{T}f\|&=&\|{\cal F} \{f\}- {\cal F}\{F_{W}S_{T}f\}\|\nonumber\\
&\leq&\|{\cal F} \{f\}-{\cal F} \{F_{W}f\}\|+\|{\cal F} \{F_{W}f\}- {\cal F} \{F_{W}S_{T}f\}\|\nonumber\\
&\leq&\varepsilon_{W}+\|F_{W}f-F_{W}S_{T}f\|\nonumber\\
&\leq&\varepsilon_{W}+\|F_{W}\|\|f-S_{T}f\|\nonumber\\
&\leq&\varepsilon_{W}+\varepsilon_{T}.
\end{eqnarray}
The last step of equation (\ref{ky1}) use the fact that $\|F_{W}\|=1$.
For $$\|f\|-\|F_{W}S_{T}f\|\leq\|f-F_{W}S_{T}f\|\leq \varepsilon_{T}+\varepsilon_{W},$$ we have
$$\|F_{W}S_{T}f\|\geq 1-\varepsilon_{T}-\varepsilon_{W}.$$
Therefore, $$\|F_{W}S_{T}\|\geq 1-\varepsilon_{T}-\varepsilon_{W}.$$ Here $\|f\|=1$ is used.
By Lemma \ref{lemadd2} and Lemma \ref{lemadd3}, we complete the proof.\end{Proof}

\section{Signal Recovery Problem}\label{S4}

Donoho and Stark \cite{DS1989} studied some examples which applied the
generalized uncertainty principle (\ref{yk4eq2}) to show something
unexpected is possible. The recovery of a signal despite significant
amounts of missing information. One example is: A signal $f \in
L^2(\R^2, \H)$ is transmitted to a receiver who knows that $f$ is
bandlimited, meaning that $f$ was synthesized using only frequencies
in a set $W \in \R^2$. Now suppose that the receiver is unable to observe
all the data of $f$, a certain subset $T$ of $\ux$-values is unobserved.
Moreover, the observed signal $f\in L^2(\R^2,\H)$ is contaminated by observational noise
$n \in L^2(\R^2,\H)$. Thus the received signal $r(\ux)$
satisfies
\begin{eqnarray}
r(\ux)=\left\{
\begin{array}{lll}
f(\ux)+n(\ux), & \ux\not\in T,\\
0,& \ux \in T,
\end{array}
\right.
\label{recover}
\end{eqnarray} where  ${\rm {supp}}{\cal F}(f) \in W$.

The receiver's aim is to reconstruct the transmitted signal $f$ from the noisy received signal $r$.
Although it may seem that information of $f$ about  $\ux \in T$ is unavailable, the uncertainty principles says that the recovery is possible
provided that $|T||W| <1$. Donoho and Stark \cite{DS1989} proved this result in the one dimensional case.
We may derive the analogue result to quaternion-valued signals.

\begin{Th}\label{yk4th2}  If $W$ and $T \in\R^2$ satisfy the condition $|T||W|<1$, then $f$ can be uniquely
reconstructed from $r$. That is, there exists a linear operator $Q$ and a constant $C$
with $C\leq (1-\sqrt{|T||W|})^{-1}$ such that
$$
\|f-Qr\|\leq C\|n\|
$$
for all $f,\, r$, and the noise $n$ obeying (\ref{recover}).
\end{Th}

\begin{Proof}

\qquad

\begin{itemize}

\item[Step 1.] We first prove that $f(\ux)$ is the unique signal which
can be recovered from the observed signal $r$. Suppose that
$f_1$ can be recovered from $r$. Let $h(\ux) :=f(\ux)-f_1(\ux)$, we have
$h(\ux)=0$, for all $\ux\notin T$. While $F_{W}f(\ux)=f(\ux)$ and
$F_{W}f_1(\ux)=f_1(\ux)$, so that $F_{W}h(\ux)=h(\ux)$. That means
$h(\ux)$ is bandlimited in $W$. Then $h(\ux)$ must be zero function on $\R^2$, otherwise it would be contradiction with Theorem \ref{yk4th1}
(since the condition $|T||W|<1$). Thus $f(\ux)=f_1(\ux)$ is unique.

\item[Step 2.]
Let $Q=(I-S_TF_W)^{-1}$. Form Lemma \ref{lemadd3}, we have
$\|F_WS_T\|_{HS}=\sqrt{|T||W|}$. Using Lemma \ref{lemadd2}, Lemma
\ref{lemadd1} and the condition $|T||W|<1$, we have
$\|S_TF_W\|\leq\|S_TF_W\|_{HS}=\|F_WS_T\|_{HS}<1$. Then $Q$ exists
because the well-known argument that the linear operator $I-L$ is
invertible if $\|L\|<1$. We also have
\begin{eqnarray}\label{mg}
\|(I-L)^{-1}\|\leq(1-\|L\|)^{-1}.
\end{eqnarray}
Since $(I-S_T)f(\ux)=(I-S_TF_W)f(\ux)$ for every bandlimited $f(\ux)$ and
\begin{eqnarray*}
f(\ux)-Qr(\ux)&=&f(\ux)-Q(I-S_T)f(\ux)-Qn(\ux)\\
&=&f(\ux)-(I-S_TF_W)^{-1}(I-S_TF_W)f(\ux)-Qn(\ux)\\
&=&0-Qn(\ux),
\end{eqnarray*}
so
\begin{eqnarray*}
\|f-Qr\|=\|Qn\|\leq\|Q\|\|n\|\leq(1-\sqrt{|T||W|})^{-1}\|n\|,
\end{eqnarray*}
Eq. (\ref{mg}) is used in the last step. This complete the proof.\end{itemize}
\end{Proof}

The operator $$Q=(I-S_{T}F_{W})^{-1}=\sum_{k=0}^{\infty}(S_{T}F_{W})^k$$
suggests an algorithm for computing $Qr$.

\begin{Th} [Algorithm for signal recovery by uncertainty principle]
Suppose that $f \in L^2(\R^2, \H)$ and $f$ is $W$-bandlimited, i.e., supp ${\cal F}\{f \}  \subseteq W \in \R^2$.
Given the received signal $r$ satisfies (\ref{recover}) with the observational noise $n \in L^2(\R^2,\H)$,
then the information of $f$ about $\ux \in T$ can be recovered by the following algorithm
\begin{eqnarray*}
s^{(0)}&=&r\\
s^{(1)}&=&r+S_TF_{W}s^{(0)}\\
s^{(2)}&=&r+S_TF_{W}s^{(1)}\\
\cdots
\end{eqnarray*}
and so on, where $S_T F_W$ are given in equation (\ref{addeqma3}), provided that $|T||W|<1$. Then $s^{(n)}\rightarrow f$ as $n\rightarrow\infty$.
\end{Th}

\begin{figure}[!]
  \centering
\includegraphics[width=10cm]{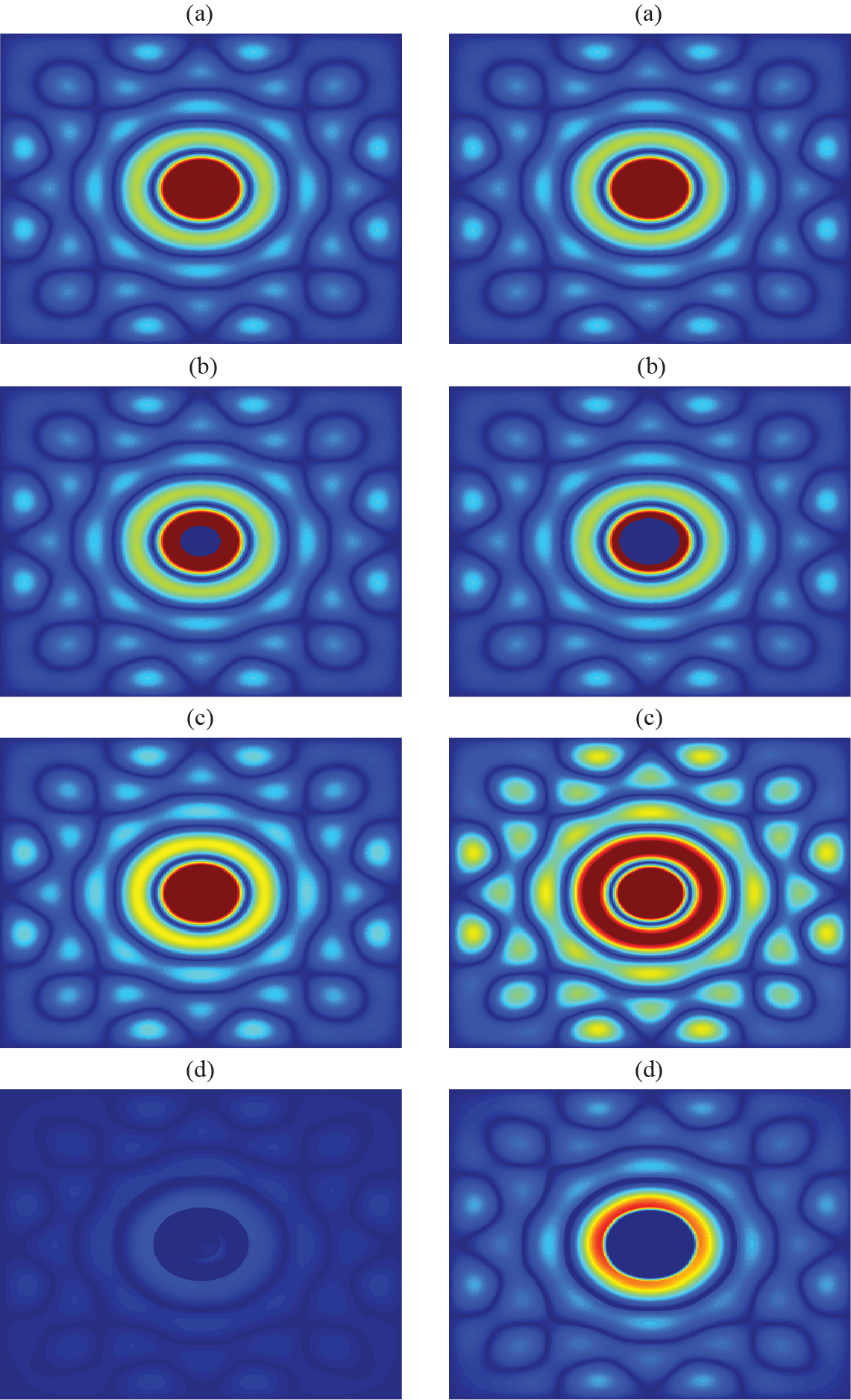}
  \caption{Signals with bandwidth $r=5$ in Example \ref{eg1}.
  (a) original signal (b) the signal with missing information, the first column is for missing $R=20$, the second column is for missing $R=30$
  (c) recovered signal (d) the difference between the original signal and recovered signal.}
 \label{fig.1D}
\end{figure}

\begin{Exa}\label{eg1}
Given the received signal $r \in L^2(\R^2, \H)$  satisfies \begin{eqnarray*}
r(\ux)=\left\{
\begin{array}{lll}
f(\ux)+n(\ux), & \ux\not\in T,\\
0,& \ux \in T,
\end{array}
\right.
\end{eqnarray*} with the observational noise $n \in L^2(\R^2,\H)$
then, for $0<r <{ 1 \over 2 \pi |T|}$, the information of $W=B({\underline{0}}, r)$-bandlimited signal $f$ about $\ux \in T$ can be recovered by the following algorithm
\begin{eqnarray*}
s^{(0)}&=&r\\
s^{(1)}&=&r+S_TF_{W}s^{(0)}\\
s^{(2)}&=&r+S_TF_{W}s^{(1)}\\
\cdots
\end{eqnarray*}
and so on, where $B(0, r)$ is the circle with center $\underline{0} \in \R^2$ and radius $r >0$ and $S_T F_W$ on $s \in L^2(\R^2, \H)$ are given by
\begin{eqnarray*}
(S_{T} F_{W}f)(\underline{x})=\chi_{T}(\ux)\int_{\R^2}s(\ut)k(\underline{t}, \ux)d\ut
\end{eqnarray*}
where
\begin{eqnarray}\label{eq1kernel}
k(\ut, \underline{x} ) = \left\{ \begin{array}{ll} \frac{1}{(2\pi)^2} \int_{0}^{2 \pi} \int_0^1 e^{-\i t_1 r \cos \theta} e^{\j (x_2-t_2)r \sin \theta} e^{\i x_1 r \cos \theta }r dr d \theta, & \ut \in T \mbox{ and }  \ux \in \R^2,\\
0, & \mbox{ otherwise.}\end{array}\right.
\end{eqnarray}
Then $s^{(n)}\rightarrow f$ as $n\rightarrow\infty$.

Here, the number $1$ in the inequality $|T||W|<1$ of Theorem
\ref{yk4th2} is corresponding to the normalized signals. In real
world, most of signals are not unit energy signals. In Fig.
\ref{fig.1D}, we construct a  simulation signal with the sustained
domain $[-20,20]$. The original signal is generated by  the inverse
Fourier transform of a rectangular function with band $[-10,10]$ (radius $r=10 \sqrt{2}$) and
the $W=10$  in Fig. \ref{fig.1D}(a). For this signal, we consider
the inequality  $TW<20$ and $W=10$. That is to say, the condition of
the limit of $T$ equals to $2$, i.e., the information missing in the
time domain is no more bigger than $[-2,2]$  in Fig.
\ref{fig.1D}(b). The signal $f(t)$ is recovered by the proposed
algorithm in Fig. \ref{fig.1D}(c). In order to get this recovered
signal, we iterate $10000$ times and show the different from the
recovered signal to original signal in Fig. \ref{fig.1D}(d). We can
find that the information is filled in the missing parts of the
signal and most of the information for the recovered signal is still
the same as the original signal. Hence, this method is effective on
this example.
\end{Exa}

\begin{Exa}
Now, if the original signal $f(\ux)$ ($\ux=x_1 \i +x_2 \j$) is bandlimited in
$W=[-\Omega, \Omega]\times[-\Omega, \Omega]$, and we would like to recover it from observed data $r$, then by applying the proposed algorithm,
\begin{eqnarray*}
s^{(0)}&=&r\\
s^{(1)}&=&r+s^{(0)}*\rm{sinc}(t_1, t_2)\\
\cdots\\
s^{(n)}&=&r+s^{(n-1)}*\rm{sinc}(t_1, t_2),
\end{eqnarray*}
where $$\rm{sinc}_{\Omega}(t_1, t_2) :=\frac{\sin(\Omega t_1)}{\pi t_1}\frac{\sin(\Omega t_2)}{\pi t_2}.$$
In fact, from (\ref{addeqma2}), we have
\begin{eqnarray*}
k(\ut, \ux)&=&\frac{1}{(2\pi)^2}\int_{W}e^{-\i t_1 \omega_1}e^{-\j t_2\omega_2}e^{\j x_2\omega_2}e^{\j x_1\omega_1}d\underline{\omega}\\
&=&\frac{1}{(2\pi)^2}\int_{[-\Omega, \Omega]}e^{-\i t_1 \omega_1} \left[\int_{[-\Omega, \Omega]}e^{-\j t_2\omega_2}e^{\j x_2\omega_2}d\omega_2 \right] e^{\j x_1\omega_1}d{\omega_1}\\
&=&\frac{1}{\pi^2}\frac{\sin \Omega(t_1-x_1)}{(t_1-x_1)}\frac{\sin \Omega(t_2-x_2)}{(t_2-x_2)}\\
&=&\frac{\sin \Omega(t_1-x_1)}{\pi(t_1-x_1)}\frac{\sin \Omega(t_2-x_2)}{\pi(t_2-x_2)}\\
&=&\rm{sinc}_{\Omega}(t_1-x_1, t_2-x_2) =\rm{sinc}_{\Omega}(x_1-t_1,
x_2-t_2).
\end{eqnarray*}
\end{Exa}
Using (\ref{addeqma3}), we obtain
\begin{eqnarray*}
P_TP_W f(\ux)&=&\int_{\R^2}f(\ut)k(\ut, \ux)d\ut\\
&=&\int_{\R^2}f(t_1, t_2)\rm{sinc}(x_1-t_1, x_2-t_2)dt_1dt_2\\
&=& f*\rm{sinc}_{\Omega}(x_1, x_2), \mbox{ }(x_1, x_2) \in T.
\end{eqnarray*}

\section{Experiments}\label{S5}

\begin{figure}[!]
  \centering
\includegraphics[width=10cm]{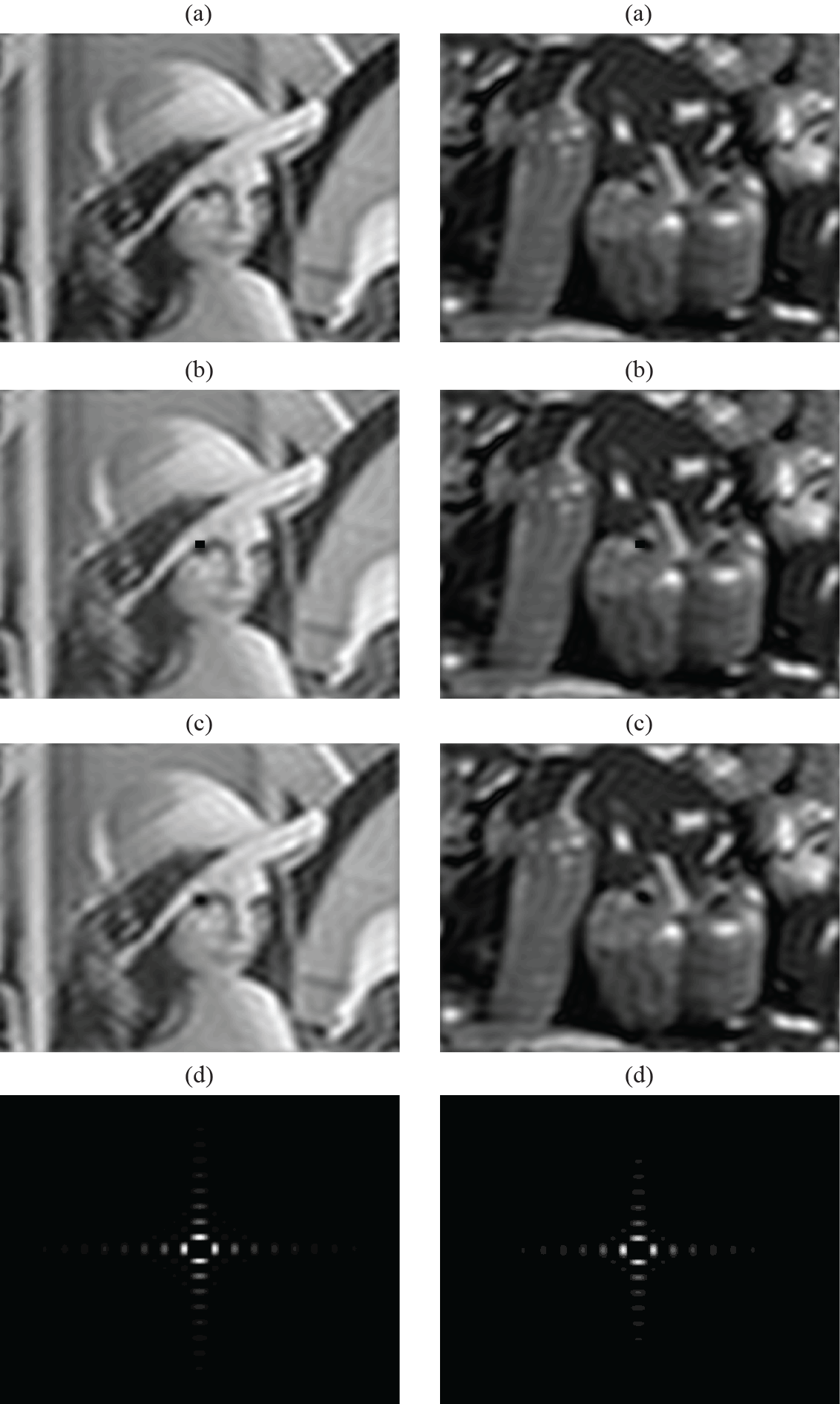}
  \caption{Example of  bandlimited Lena and Chillies with $W=80$.
  (a) original  bandlimited images (b) the  bandlimited images with missing information
  (c) recovered  bandlimited images (d) the difference between the original  bandlimited images and recovered  bandlimited images.}
 \label{fig.2D1}
\end{figure}

\begin{figure}[!]
  \centering
\includegraphics[width=10cm]{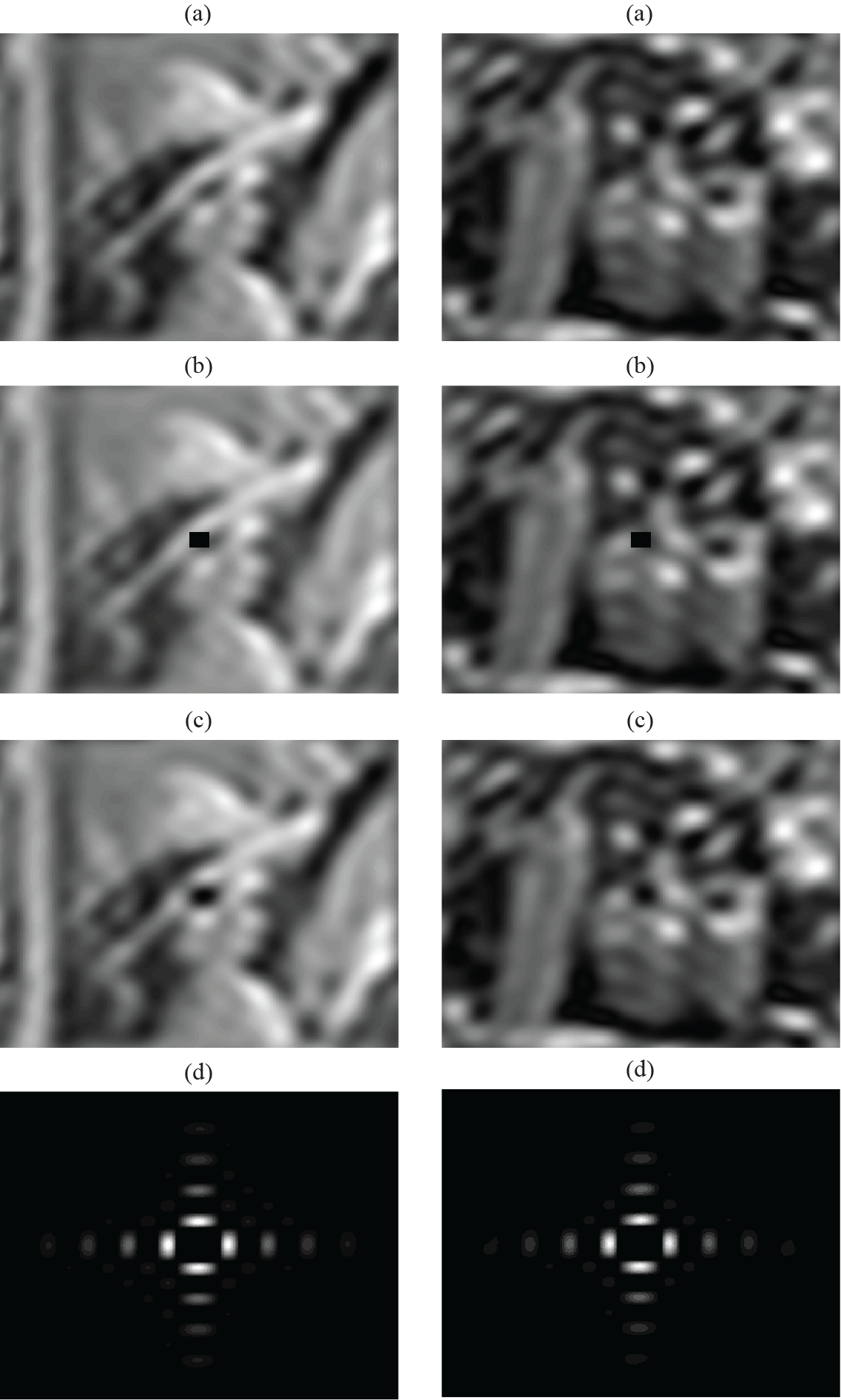}
  \caption{Example of  bandlimited Lena and Chillies with $W=40$.
  (a) original  bandlimited images (b) the  bandlimited images with missing information
  (c) recovered  bandlimited images (d) the difference between the original  bandlimited images and recovered  bandlimited images.}
 \label{fig.2D2}
\end{figure}

In this section, two experiments are carried out to test the performance of the proposed algorithm as example.
The reconstruction results are shown in Figs. \ref{fig.2D1}-\ref{fig.2D2}.

Two bandlimited images are constructed by Lena and Chillies, which size is size $400 \times 400$.
Two bandlimited Lena and Chillies are constructed with bandwidth $W=80$ and $W=40$, which are the original images and shown in Fig. \ref{fig.2D1}(a) and Fig. \ref{fig.2D2}(a).
Here, the condition of $|T||W|<1$ in Theorem \ref{yk4th2} becomes $TW<400$, since the whole domain is $400 \times 400$.
That means for Lena and Chillies with  bandwidth $W=80$, $T$ cannot exceed $5$, i.e., the information missing block in this image is less than  $5 \times 5$.
As for the Lena and Chillies with  bandwidth $W=40$, this missing region must be less than $10\times 10$.
In Fig. \ref{fig.2D1}(b) and Fig. \ref{fig.2D2}(b), the two missing information Lena and Chillies are shown,
where the two black rectangular block are the regions we generated with no information.
In Fig. \ref{fig.2D1}(c) and Fig. \ref{fig.2D2}(c), the two recovered images are shown.
And the errors between the recovered images and the original Lena and Chillies images are shown in Fig. \ref{fig.2D1}(d) and Fig. \ref{fig.2D2}(d).
The error for the Lena and Chillies with bandwidth $W=80$ is smaller than the Lena and Chillies with  bandwidth $W=40$.
From the two error images we can found that, except the center of  image, the difference for original images and recovered images is black.
That is to say, for these regions with no information missing, there is little changes.
For the regions with information missing, there is some information be filled.
This makes sense of the proposed method.

\section{Conclusions and Discussions}\label{S6}
In this paper, we have proposed the Quaternion Fourier transform (QFT) for signal recovery problems.
The mathematical definitions of QFT for measurable sets are first presented.
Then two crucial operators namely space-limiting and frequency-limiting operators are discussed.
Applying their properties, the uncertainty principle for measurable sets associated with QFT are given.
Finally, the image representation capabilities are discussed by experiments on real images.
Experimental results have demonstrated that the proposed algorithms have achieved promising results.
As future works, we will apply the proposed QFT in a variety of applications, such as color image retrieval and
color image watermarking. The generalized integral transformations namely Quaternion fractional Fourier transform and
Quaternion linear canonical transform will also be considered in the upcoming paper.

By the non-commutation for quaternions, there are various kinds
of quaternion Fourier transforms (QFTs). For example, the left-sided, the
right-sided (this case in our paper) and two-sided QFTs. The theory
about the left-sided case is parallel to the right-sided. For the
two-sided quaternion Fourier transform, the present methods in the cannot be used.
The reason is as follows.

We first recall the definition of two-sided quaternion Fourier
transform as:
\begin{eqnarray*}\label{2QFT} {\cal F}_2 \{f\}(\uxi)
:=\frac{1}{2\pi}\int_{\R^2}e^{-\i x_1\xi_1}f(\ux)e^{-\j
x_2\xi_2}d\ux
\end{eqnarray*}
and if in addition, ${\cal F}_2 \{f\} \in L^1\bigcap L^2(\R^2,\H)$,
function $f$ can be recovered by its QFT \cite{CK2016, HK2016} as
$$f(\ux)=\frac{1}{2\pi}\int_{\R^2}e^{\i x_1\xi_1}{\cal F}_2 \{f\}(\uxi)e^{\j x_2\xi_2}d\uxi.$$

Therefore equation (\ref{addeqma1}) becomes
\begin{eqnarray*}
(F_{W} S_{T}f)(\underline{x})&=&\frac{1}{2\pi}\int_{W}e^{\i x_1\omega_1}{\cal{F}_2}\{S_Tf\}(\underline{\omega})e^{\j x_2\omega_2}d\underline{\omega}, \quad \ux \in \R^2\nonumber\\
&=&\frac{1}{(2\pi)^2}\int_{W}\left(\int_{T}e^{\i x_1\omega_1}e^{-\i t_1\omega_1}f(\ut)e^{-\j t_2\omega_2}d\ut\right)e^{\j x_2\omega_2}e^{\i x_1\omega_1}d\underline{\omega}\nonumber\\
&=&\frac{1}{(2\pi)^2}\int_{T}\int_{W}e^{\i (x_1-t_1)\omega_1}f(\ut)e^{-\j (t_2-x_2)\omega_2}d\underline{\omega}d\ut\nonumber.\\
\end{eqnarray*}
By the non-commutative of quaternions, if $f(\ut)$ is
quaternion valued, we cannot take $f(\ut)$ out, so we cannot define
the Hilbert-Schmidt norm of $F_{W} S_{T}$. Alternative method will be considered in future studies.

\section*{Acknowledgment}
This work is supported by the National Natural Science Foundation of China (11401606 and 11501015), University of Macau MYRG2015-00058-L2-FST and the Macao Science and Technology
Development Fund (FDCT/099/2012/A3 and FDCT/031/2016/A1).


\end{document}